\documentclass[aps,pra,twocolumn,floatfix,showpacs,superscriptaddress]{revtex4}

\usepackage{color}
\usepackage{graphicx}
\usepackage{bbm}
\usepackage{amsmath,amssymb}

\newcommand{\be}{\begin{equation}}
\newcommand{\beq}{\begin{equation}}
\newcommand{\ee}{\end{equation}}
\newcommand{\bea}{\begin{eqnarray}}
\newcommand{\eea}{\end{eqnarray}}
\newcommand{\ba}{\begin{array}}
\newcommand{\ea}{\end{array}}

\renewcommand{\vr} {{\bf r}}

\newcommand{\vk} {{\bf k}}

\begin{document}
\title{On the validity of power functionals for the homogeneous electron gas \\ in reduced-density-matrix-functional theory}
\author{A. Putaja}
\affiliation{Nanoscience Center, Department of Physics, University of
  Jyv\"askyl\"a, FI-40014 Jyv\"askyl\"a, Finland}
\affiliation{Department of Physics, Tampere University of Technology, FI-33101 Tampere, Finland}
\author{F. G. Eich}
\affiliation{Max Planck Institute for the Structure and Dynamics of Matter, Luruper Chaussee 149, 22761 Hamburg, Germany}
\affiliation{Department of Physics, University of Missouri-Columbia, Columbia, Missouri 65211, USA}
\author{T. Baldsiefen}
\affiliation{Max-Planck-Institut f\"ur Mikrostrukturphysik, Weinberg 2, D-06120 Halle, Germany}
\affiliation{Jenoptik Optical Systems GmbH, Jena, Germany}
\author{E. R{\"a}s{\"a}nen}
\email[Electronic address:\;]{esa.rasanen@tut.fi}
\affiliation{Department of Physics, Tampere University of Technology, FI-33101 Tampere, Finland}
\affiliation{Nanoscience Center, Department of Physics, University of
  Jyv\"askyl\"a, FI-40014 Jyv\"askyl\"a, Finland}

\date{\today}

\begin{abstract}
Physically valid and numerically efficient approximations for the exchange and
correlation energy are critical for reduced-density-matrix-functional theory to become a widely used method in electronic structure calculations. Here we examine the physical limits of power functionals of the form $f(n,n')=(n n')^\alpha$ for the scaling function in the exchange-correlation energy. To this end we obtain numerically the minimizing momentum distributions for the three- and two-dimensional homogeneous electron gas, respectively. In particular, we examine the limiting values for the power $\alpha$ to yield physically sound solutions that satisfy the Lieb-Oxford lower bound for the exchange-correlation energy and exclude pinned states with the condition $n(\vk)<1$ for all wave vectors $\vk$. The results refine the constraints previously obtained from trial momentum distributions. We also compute the values for $\alpha$ that yield the exact correlation energy and its kinetic part for both the three- and two-dimensional electron gas. In both systems, narrow regimes of validity and accuracy are found at $\alpha\gtrsim 0.6$ and at $r_s\gtrsim 10$ for the density parameter, corresponding to relatively low densities.
\end{abstract}

 
\maketitle

\section{Introduction}

Reduced-density-matrix-functional theory~\cite{lowdin,gilbert} 
(RDMFT) has attracted interest and popularity as an
alternative to density-functional theory~\cite{dft} (DFT) 
to deal with complicated many-particle problems.
In contrast with the one-body density in DFT,
the key quantity in RDMFT is the one-body reduced density matrix (1-RDM),
which provides the exact kinetic energy. 
It is thus evident that RDMFT can outperform
DFT in strongly correlated systems~\cite{corre1,corre2,corre3}, and
recent extensions and investigations include, e.g., 
finite temperatures~\cite{finitetemp}, excitation 
energies~\cite{excitations}, and Mott insulators~\cite{mott1,mott2}. 
However, so far only a few energy functionals 
of the 1-RDM have been developed, and their practical 
applicability is still partly unknown.

The so-called power functionals~\cite{sharma} of RDMFT
have been applied in three-dimensional (3D)
systems during the past few years~\cite{nek1,nek2,mott2}. 
In this functional the scaling factor for the exchange-correlation ($xc$) 
energy $E_{xc}$ has a form $f(n,n')=(n n')^\alpha$,
where $1/2\leq\alpha\leq 1$ can be viewed as a parameter
interpolating between the Hartree-Fock (HF) ($\alpha=1$)
and M\"uller~\cite{muller} ($\alpha=0.5$) approximations.
The optimal values for $\alpha$ have been found to
vary between $0.525$ (stretched $H_2$) and $0.65$ (solids).
The best overall fit for the 3D homogeneous 
electron gas (3DEG) has been obtained with $\alpha=0.55\ldots 0.58$ 
\cite{nek2}. In the two-dimensional (2D) framework,
the applications are more scarce. However,
Harju and T\"ol\"o~\cite{harju} have found reasonable
results for 2D quantum Hall droplets at high magnetic 
fields with $\alpha\sim 0.65\ldots 0.7$. 

Power functionals are also subject to physical constraints of
RDMFT~\cite{levy}. In the case of the 3DEG, strict constraints
regarding the solution of the Euler-Lagrange equation and
the Lieb-Oxford (LO) bound~\cite{lo} have been studied by 
Cioslowski and Pernal~\cite{kasia,kasia2}. A similar study
on the 2D homogeneous electron gas (2DEG) has been recently
carried out by some of the present authors with a 
particular attention on accessible densities~\cite{antti}. 
However, both of these studies have resorted to analysis of 
trial momentum distributions with uncertainty of their accuracy
in comparison with numerically exact results.

In this work we numerically obtain the minimizing
momentum distributions for the power functional in both 
the 3DEG and 2DEG. The resulting range of validity for $\alpha$
at various densities is then compared to the
results obtained previously with trial momentum 
distributions in 3D~\cite{kasia} and 2D~\cite{antti}, 
respectively. In both cases we begin with the general
constraints on $n(\vk)$ and on the densities $\rho$. We then proceed 
with a closer analysis of (i) the LO bound~\cite{lo} with its recently 
suggested tighter forms~\cite{ourbound}, and (ii) the {\em exclusion} 
of pinned states with $n(\vk)=1$~\cite{kimball,klaas1}.

We find that in both 3D and 2D the power functionals have
a rather limited range of validity, and only for relatively
low densities, although the numerical
solutions extend the range in comparison with the previously 
used trial momentum distributions. We also calculate numerically 
the exact correlation energies and their kinetic contributions
as a function of $\alpha$. The exact solutions coincide only partly, 
if at all, with the regimes of validity.

\section{Homogeneous electron gas}\label{homo}

For the homogeneous electron gas (EG), here generally in 3D or 2D, 
we can consider a positive background charge 
compensating for the electrostatic (Hartree) energy,
so that the total energy consists of the kinetic and xc 
components alone, i.e.,
\be
E^{\rm EG}_{\rm tot}[\gamma] = T[\gamma] + E_{xc}[\gamma].
\ee
In RDMFT, we can express the kinetic and xc energies
(in Hartree atomic units) as
\be
T[\gamma] = -\frac{1}{2}\sum_{\sigma=\uparrow,\downarrow}\sum_p n_\sigma(\vk_p)\int d\vr \,\varphi^*_{p\sigma}(\vr) \nabla^2 \varphi_{p\sigma}(\vr)
\label{T2}
\ee
and
\bea\label{exc2}
E_{xc}[\gamma]  &    =   &  -\frac{1}{2}\sum_{\sigma=\uparrow,\downarrow}\sum_{p,q}^\infty  \int d\vr \int d\vr' f\left(n_\sigma(\vk_p),n_\sigma(\vk_q)\right) \nonumber \\
       & \times & \frac{\varphi^*_{p\sigma}(\vr)\varphi^*_{q\sigma}(\vr')\varphi_{q\sigma}(\vr)\varphi_{p\sigma}(\vr')}{|\vr-\vr'|}.
\eea
Here $\vk_p$ is the wave vector of the $p$th spin-dependent natural
orbital, which is a plane wave.
Now, in the thermodynamic limit the summation over plane waves is placed by
a momentum-space integration and, hence, we can express the total energy as a
functional of the momentum distribution~\cite{kasia,antti}.
Using the normalization constraint and the variational Euler-Lagrange
equation finally leads to
\be
\frac{1}{2}|\vk|^2 - \frac{1}{2\pi^{d-1}}\int d\vk' \, \frac{\frac{\partial}{\partial n(\vk)}f\left(n(\vk),n(\vk')\right)}{|\vk-\vk'|^{d-1}} = \mu,
\label{n}
\ee
where $\mu$ is the Lagrange multiplier, and $d=3$ in 3D and $d=2$ in 2D.
Note that the spin index $\sigma$ has been omitted here, and in the
following we consistently refer to quantities per particle
with spin $\sigma$. It is important to appreciate that the Euler-Lagrange equation only holds for all $\vk$ if the
minimizing momentum distribution has no pinned states, i.e., $n(\vk)\neq 1$
and $n(\vk) \neq 0$ for all wave vectors $\vk$. 

The key quantity in the expressions above is the function 
$f\left(n(\vk),n(\vk')\right)$ used in RDMFT as a scaling factor
to take into account electron--electron correlations beyond
the mean-field or HF level. Our main focus here is on the
generic power functional, 
\be
f\left(n(\vk),n(\vk')\right) = \left(n(\vk) n(\vk')\right)^{\beta/2} = \left(n(\vk) n(\vk')\right)^{\alpha}.
\label{el}
\ee
Here $\beta=2$ ($\alpha=1$) and $\beta=1$ ($\alpha=1/2$) correspond to the
HF and the M\"uller functional~\cite{muller}, respectively. The parameter
$\beta$ is used below instead of $\alpha$ in order to ease the comparison with
Refs.~\cite{kasia} (3D) and \cite{antti} (2D).

\subsection{Three-dimensional case}\label{3d}

\subsubsection{General constraints}

Following Ref.~\cite{kasia}, we first briefly review some fundamental constraints 
for $\beta$ in 3D. We restrict ourselves to fully variational solutions 
of Eq.~(\ref{n}); in that case the solutions scale with the electronic density $\rho$ as
\be
n(\vk)= \rho^{1/(3\beta-2)}\eta\left(\rho^{\frac{1-\beta}{3\beta-2}}\vk\right),
\label{ansatz}
\ee
where $\eta(x)$ is independent of the density. 
Using the constraint $0\leq n(\vk) \leq 1$ with a homogeneous scaling requirement for $f\left(n(\vk),n(\vk')\right)$ leads to 
a criterion $\beta>2/3$. Further, physical constraints of positive kinetic-energy density 
$t$ and nonpositive xc energy density $\epsilon_{xc}$ (defined per volume in this work) lead to $\beta<4/3$. 

Only a finite range of densities is allowed in the obtained range,
$2/3<\beta<4/3$. First, $n(\vk) \leq 1$ yields
a criterion 
\be
\rho\leq \eta_{\rm max}^{2-3\beta}, 
\label{rho1}
\ee
where $\eta_{\rm max}$ is the maximum value of $\eta(x)$. 
Secondly, the Lieb-Oxford (LO) lower bound~\cite{lo} for $E_{xc}$ (of spin-unpolarized gas) yields
\be
\epsilon_{xc}\geq -C_{\rm 3D}(2\rho)^{1/3},
\label{lob}
\ee
where the factor of two results from per-spin notation (see above).
Here $C_{\rm 3D}=C_{\rm 3D}^{\rm LO}=1.68$ according to the rigorous LO bound~\cite{lo}, and $C_{\rm 3D}=C_{\rm 3D}^{\rm RPCP}=1.44$
according to a tighter, nonrigorous bound in Ref.~\cite{ourbound}; see also Refs.~\cite{lieb2015,paola2016} for recent analysis. We consider both of these bounds in the following. Generally, the bound in Eq.~(\ref{lob}) holds only for densities
\be
\rho\geq \left[4(3\beta-2)^{-3}(-A_\epsilon)^{3}C_{\rm 3D}^{-3}\right]^{\frac{3\beta-2}{4-3\beta}},
\label{rho2}
\ee
where
\bea
\label{aepsilon}
 A_{\epsilon} & = & \frac{1}{16\pi^3}\int d\vk\,\eta(\vk)|\vk|^2 \nonumber \\
                       & - & \frac{1}{32\pi^5}\int d\vk \int d\vk' \, \frac{f\left(\eta(\vk),\eta(\vk')\right)}{|\vk-\vk'|^{2}}.
\eea

To summarize the present section, there is a general constraint $2/3<\beta<4/3$ in the 3D 
power functional. In addition, the allowed densities are restricted by 
Eqs.~(\ref{rho1}) and (\ref{rho2}). In the following we examine how these constraints change as we consider
either a trial momentum distribution of Ref.~\cite{kasia}, or a numerical one that 
minimizes the energy functional.

\subsubsection{Trial momentum distribution}

Cioslowski and Pernal~\cite{kasia} considered a parametrized {\em trial} function 
for $\eta$ similar to that of the M\"uller functional,
\be
{\bar \eta}(\vk) = D(\beta,\zeta)(1+\zeta |\vk|^2)^{-4/\beta},
\label{momentumpower}
\ee
which is exact for $\beta=1$.
Here $D$ is a normalization constraint, and $\zeta$ is solved such that the
total energy density $\epsilon = A_\epsilon \, \rho^{(2\beta-2)/(3\beta-2)}$
is minimized. This corresponds to the minimization of the integral $A_\epsilon$ in Eq.~(\ref{aepsilon}).
Now, using the condition $n(\vk) \leq 1$ yields
\begin{equation}
\rho \leq \left[2^{8/\beta}\pi\,Q(\beta)\zeta_{m}^{3/2}\right]^{2-3\beta},
\label{3Drho1}
\end{equation}
where
\begin{equation}
\zeta_m(\beta) = \left[\frac{3\pi^{1-\beta}}{(16\beta^{-1}-10)(4-3\beta)Q(\beta)^{\beta}}\right]^{\frac{2}{3\beta-2}},
\end{equation}
and
\begin{equation}
Q(\beta)= \frac{(8\beta^{-1}-1)(8\beta^{-1}-3)\Gamma(4/\beta)^{2}}{\Gamma(8/\beta)}.
\end{equation}
The lower bound of $\rho$ is obtained from Eq.~(\ref{rho2}) by minimizing $A_\epsilon$ with the trial wave function~\cite{kasia}. Combining the results implies $\beta \geq 1.113$ with $C_{\rm 3D}^{\rm LO}$ (Ref.~\cite{lo}) and $\beta \geq 1.168$ with $C_{\rm 3D}^{\rm RPCP}$ (Ref.~\cite{ourbound}). In both cases, the upper limit $\beta<4/3$ naturally applies (see above). 

\subsection{Two-dimensional case}\label{2d}

\subsubsection{General constraints}

The 2D case has been considered in detail in Ref.~\cite{antti}.
Here we summarize only the main findings: The general constraints
implied by the solution of the Euler-Lagrange equation \eqref{el},
the homogeneous scaling of $f\left(n(\vk),n(\vk')\right)$, and
physical $\epsilon_{xc}$ and $t$ lead to $1/2<\beta<3/2$. 

The main difference between 3D and 2D results arises from
dimension-dependent scaling relations.
In 2D, $n(\vk)$ scales with the density as 
\be
n(\vk)= \rho^{1/(2\beta-1)}\eta\left(\rho^{\frac{1-\beta}{2\beta-1}}\vk\right).
\label{ansatz2d}
\ee
Now, $n(\vk) \leq 1$ leads to 
\be
\rho\leq \eta_{\rm max}^{1-2\beta}, 
\label{2Drho1}
\ee
where $\eta_{\rm max}$ is the maximum value of $\eta(x)$. In addition,
the LO bound has a different scaling and constant in 2D. 
The existence of the lower bound for $\epsilon_{xc}$ has been rigorously
proved~\cite{lsy}, and the tightest form for this bound has been suggested
in Ref.~\cite{ourbound}. In summary, in 2D we have
\be
\epsilon_{xc}\geq -C_{\rm 2D}(2\rho)^{1/2}
\ee
with $C_{\rm 2D}=1.96$. The bounds holds for densities
\be
  \rho\geq \left[2(2\beta-1)^{-2}{I}_\epsilon^{2}
  C_{\rm 2D}^{-2}\right]^{\frac{2\beta-1}{3-2\beta}},
\label{2Drho2}
\ee
where
\bea
\label{iepsilon}
 I_\epsilon & = & \frac{1}{8\pi^2}\int d\vk\,\eta(\vk)|\vk|^2 \nonumber \\
 & - & \frac{1}{16\pi^3}\int d\vk \int d\vk' \,
 \frac{f\left(\eta(\vk),\eta(\vk')\right)}{|\vk-\vk'|}.
\eea

\subsubsection{Trial momentum distribution}

In Ref.~\cite{antti} a parametrized trial function for $\eta$ in 2D was suggested:
\be
{\bar \eta}(\vk) = D(\beta,\zeta)(1+\zeta |\vk|^2)^{-3/\beta},
\ee
which is exact for $\beta=1$. The strategy to generalize the ansatz for arbitrary
$\beta$ is similar to the 3D case, i.e., the total energy density
$\epsilon = I_\epsilon \, \rho^{(2\beta-2)/(2\beta-1)}$ is minimized
through the integral $I_\epsilon$ in Eq.~(\ref{iepsilon}). In contrast to the 3D
case, the integral in $I_\epsilon$ requires a numerical solution.
\footnote{See Ref.~\cite{antti} for details.}

In 2D, using the condition $n(\vk) \leq 1$ yields the upper bound
for the density, i.e.,
\be
\rho \leq D^{1-2\beta} = \left[4\pi\zeta_m (3\beta^{-1}-1)\right]^{1-2\beta},
\ee
where $\zeta_m$ is the value for $\zeta$ in the trial wave function
that minimizes the integral $I_\epsilon$. The lower bound for the
density is obtained from Eq.~(\ref{iepsilon}) through minimization.
In Ref.\ \cite{antti} this was shown to yield the condition 
$1.28 \leq \beta\leq 3/2$.

\subsection{Numerical momentum distribution}

In order to assess the quality of the analytical momentum distributions
we calculate the momentum distribution numerically by minimizing the
energy functional under the $N$-representability constraints.
In particular, we compute the limiting values for the density parameter
$r^{\rm 3D}_s=\left[3/(8\pi\rho)\right]^{1/3}$ (3D) and 
$r^{\rm 2D}_s=(2\pi\rho)^{-1/2}$ (2D), for which the minimizing
momentum distribution has border minima, i.e., occupation numbers pinned
to $n(\vk)=1$. Furthermore, the LO bound is considered with $\epsilon_{xc}$
calculated with the numerically obtained momentum distribution.

Assuming that the occupation numbers are spherically symmetric, we can
discretize the momentum space into spherical volume elements ${\Omega_j}$, i.e.,
shells (3D) or rings (2D) with thickness $\delta k$.
Averaging the occupation numbers over the volume elements ${\Omega_j}$, i.e., 
\begin{align}
  n_{j \sigma} & = \int_{\Omega_j}d\vk n_{\sigma}(k) , \label{n_k_discretize}
\end{align}
and defining the integral weights
\begin{align}
  \mathrm{DWI}_j & = 
  \frac{1}{(2 \pi)^d} \int_{\Omega_j} \!\! d\vk , \label{DWI_def} \\
  \mathrm{DKI}_j & =
  \frac{1}{2 (2 \pi)^d} \int_{\Omega_j} \!\! d\vk  k^2, \label{DKI_def} \\
  \mathrm{DXI}_{jk} & =
  \frac{d-1}{(2\pi)^{2d -1}}
  \iint_{\Omega_j,\Omega_k} \!\!\!\!\!\!\!\! d\vk_1 d\vk_2
  \frac{1}{|\vk_1-\vk_2|^{d-1}} , \label{DXI_def}
\end{align}
the discretized version of the energy functional reads
\begin{align}
  E[\{n_j\}] & = \sum_{j\sigma} n_{j \sigma} \mathrm{DKI}_j - \mu \sum_{j \sigma} n_{j \sigma} \mathrm{DWI}_j \nonumber \\
  & - \frac{1}{2} \sum_{jk\sigma} f(n_{j\sigma};n_{k\sigma}) \mathrm{DXI}_{jk} . \label{E_RDMFT_disc}
\end{align}
We note that all integral weights, Eqs.\ \eqref{DWI_def}--\eqref{DXI_def},
can be solved analytically~\footnote{In this context ``analytical'' means that
the integrals can be expressed in terms of special functions,
e.g., hypergeometric functions for the 2D exchange integral.}. Furthermore,
the procedure of taking the occupation numbers constant in spherical
volume elements implies that we always treat the wave vector $\vk$ as a
continuous variable. Accordingly, all calculations are done 
in the thermodynamic limit,
which means that the volume, $V$, and the number of particles, $N$, tend
to infinity, while the ratio $\rho = N / V$ remains constant. 
Hence, the computed
total energies are variational--meaning upper bounds--to the
true ground-state energy for a given functional.
The size of the spherical volume elements determines
the quality of this upper bound for a given functional.

The minimization of the functional Eq.~\eqref{E_RDMFT_disc} is a
high-dimensional non-linear optimization problem in
terms of the occupation numbers $\{n_j\}$. The chemical
potential ${\mu}$ is a Lagrangian multiplier ensuring that
the minimum configuration $\{n_j\}_0$ is normalized to
the required density ${\rho=\sum_{j \sigma} n_{j \sigma} \mathrm{DWI}_i}$.
The minimization is carried out using the scheme proposed
in Ref~\cite{baldsiefen}, which employs a fictitious non-interacting
electron gas at finite temperature in order to constrain the
occupation numbers to be ${n_j\in [0,1]}$. The momentum distribution
is linearly sampled by $N$ volume elements for $k \in [0, k_c]$ and
the tail of the momentum distribution, from $k \in [k_c, 100 k_c]$,
is logarithmically sampled by $N$ volume elements~\footnote{We have checked
the convergence of the presented results with respect to the discretization
parameters $N$ and $k_c$. The given results are obtained with $k_c = 2 k_f$
and $N=400$.}.

\section{Results}

\begin{figure}
\includegraphics[width=\columnwidth]{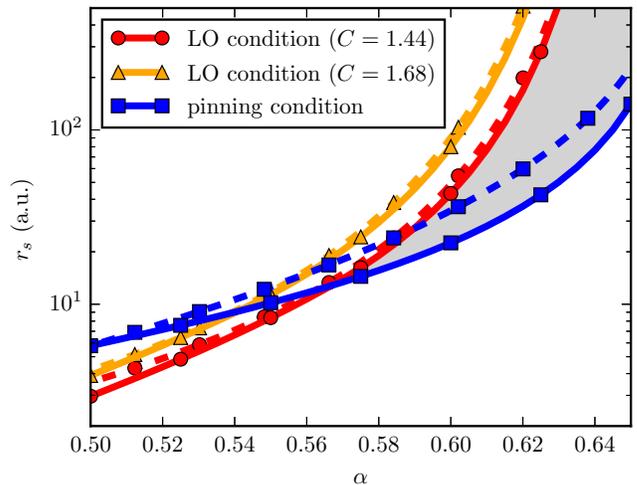}
\caption{(Color online) Density-dependent bounds on the power functional 
for the the 3DEG. The dashed lines refer to the results from the variational ansatz and the solid lines
correspond to numerical results. The blue (squares) lines separate
the regions with pinned states (below the curves) and without pinned states (above the curves).
The red (circles) and orange (triangles) lines denote boundary of the regions where
the LO bound is obeyed (below the curves) and the regions where it is violated (above the curves)
for different values of the constant $C$.\label{fig1}}
\end{figure}

\begin{figure}

\includegraphics[width=\columnwidth]{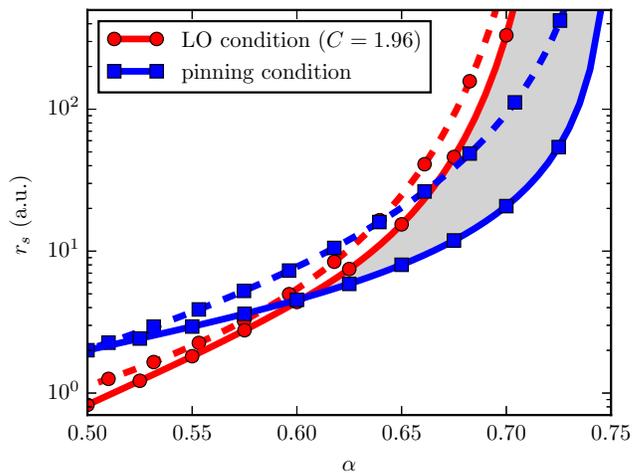}
\caption{(Color online) Same as Fig.\ \ref{fig1} but for the 2DEG.
Here we only show the result for the tight LO-type bound in 2D ($C=1.96$) according to Ref.~\cite{ourbound}. \label{fig2}}
\end{figure}

In Figs.~\ref{fig1} and \ref{fig2} we show the results for the 3DEG and
2DEG, respectively. The solid lines correspond
to critical densities obtained from numerical momentum distributions
and the dashed lines refer to the variational ansatz for the momentum
distributions discussed in Secs.~\ref{3d} (3DEG) and \ref{2d} (2DEG).
We plot the critical densities, characterized by the Wigner-Seitz radius $r_s$,
as function of the exponent $\alpha = \beta / 2$. 

The two upper (red and orange) pairs of curves in Fig.~\ref{fig1} represent the critical densities for which the 
LO bound holds as an equality. The orange line corresponds to the original LO bound ($C=1.68$)
and the red curve to the tighter LO bound ($C=1.44$) proposed in
Ref.~\cite{ourbound}. In the region above the curves the LO bound is
violated. Naturally, the tighter LO bound ($C=1.44$) leads to a
smaller region of validity. The numerical results yield
a slightly smaller region of validity than the ansatz. This means that $E_{xc}$ 
computed from the variational ansatz is larger and therefore violates
the LO bound -- which is a lower bound -- for lower densities (or higher $r_s$).
The fact that the LO curves for the ansatz and the numerical
momentum distributions are very close to each other for the 3DEG indicates that the 
variational ansatz is very accurate -- at least in the region close to the critical
density determined by the LO bound. 

The lower curves (blue) in Fig.~\ref{fig1} indicate the
critical densities at which the momentum distribution acquires pinned states, i.e., 
above the curve we have $n(\vk) < 1$ for all $\vk$ and below we have $n(\vk)=1$ for some $\vk$.
Note that the ansatz for the momentum distribution breaks down below
the dashed, blue curve. The numerical results for the momentum distribution
below the solid, blue curve represent boundary minima of the RDMFT energy functional,
which means that the Euler-Lagrange equation does not hold for wave vectors
$\vk$ with $n(\vk)=1$. We see that the numerical momentum distributions
increase the region of unpinned $n(\vk)$. Since the ansatz becomes exact for
$\alpha=0.5$ (or $\beta = 1$) the solid and the dashed blue curves coincide at
this value. However, the curves for the LO bound do not coincide at $\alpha=0.5$.
Most likely this is due to the fact that the ansatz does not
provide valid results below the pinning $r_s$. Hence, the
$xc$ energies for the LO bound in the region below the dashed, blue curve
are computed with momentum distributions that violate the Pauli constraint
$0 \leq n(\vk)\leq 1$.

By comparing Figs.~\ref{fig1} and \ref{fig2} we can see that the
results from the variational ansatz are closer to the numerical results
in the 3DEG than in the 2DEG. This indicates that the variational ansatz 
for the 3DEG is closer to the true momentum distribution than the variational 
ansatz for the 2DEG.

\begin{figure}
\includegraphics[width=\columnwidth]{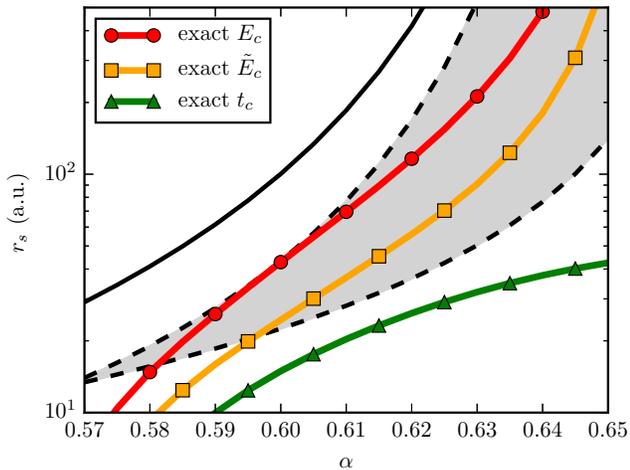}
\caption{(Color online) Exact $E_c$ (red, circles), $\tilde{E}_c$ (orange, squares), and $t_c$
(green, triangles) for the 3DEG. Within the shaded region the tight LO bound ($C=1.44$) is satisfied
and the momentum distribution is strictly smaller than one. The solid black line
denotes the violation of the LO bound when the kinetic contribution
to the correlations energy taken into account. \label{fig3}}
\end{figure}

\begin{figure}
\includegraphics[width=\columnwidth]{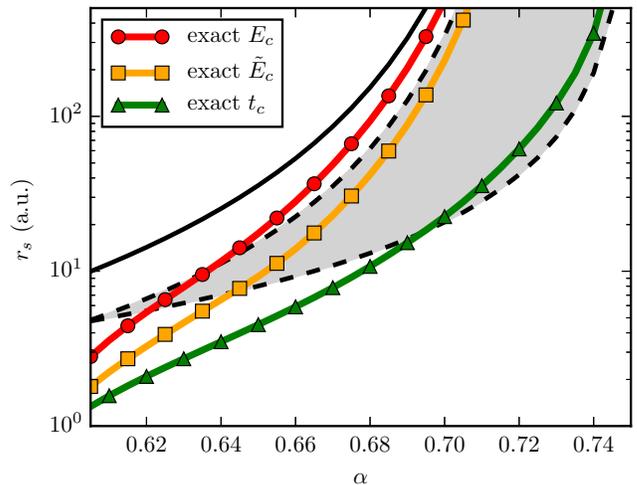}
\caption{(Color online) Same as Fig.\ \ref{fig3} but for the 2DEG.
In the shaded region the tight LO bound ($C=1.96$) is satisfied
and the momentum distribution is not pinned, i.e., $n(k)<1$. The solid black line
denotes the violation of the LO bound when the kinetic contribution
to the correlations energy taken into account.\label{fig4}}
\end{figure}

Next we consider the correlation energies $E_c$ produced by the
power functional. We remind that $E_c$ for the EG is
known exactly from quantum Monte Carlo simulations for the
3DEG \cite{OB} and the 2DEG \cite{AMGB}. The total energy per
particle of the EG is usually written as
\be
  \epsilon(r_s) = \frac{A}{r_s^2} + \frac{B}{r_s} + e_c(r_s) ~, \label{eHeg}
\ee
where $A$ and $B$ are well-known constants for the HF energy of the EG~\cite{VignaleBook}.
The correlation energy, as implicitly defined in Eq.~(\ref{eHeg}),
contains a kinetic energy contribution $t_c$, or equivalently,
we can decompose
\be
  \epsilon_c(r_s) = t_c(r_s) + \tilde{\epsilon}_c(r_s) ~, \label{ec_decomposition}
\ee
where $t_c$ is explicitly given by
\be
  t_c = \frac{1}{n} \frac{1}{(2 \pi)^d} \int \!\! d\vk k^2
  \Big[n(k) - n_0(k) \Big] ~. \label{tcDef}
\ee
Here $n(k)$ is the momentum distribution
of the interacting EG and $n_0(k)$ is the momentum distribution
of the non-interacting EG, i.e., the Fermi step function.
By using scaling relations, $\tilde{e}_c$ and $t_c$ can be obtained
from the parameterizations of $e_c$~\cite{Takada}.

In Figs.~\ref{fig3} and \ref{fig4} we show the densities for which
the total correlation energy $E_c$, the correlation energy without
kinetic contribution $\tilde{E}_c$ and the kinetic correlation energy $t_c$ are
{\em exact} as a function of $\alpha$. For the exact
functional all three quantities would be exactly reproduced. However, we can see that
for the power functional the three quantities are exact for {\em different
densities} at the same power $\alpha$. This demonstrates the fact that even
if the correlation energy is exact for a given density, the minimizing momentum
distribution is not the exact momentum distribution.

We point out that for the 3DEG the kinetic contribution to the correlation energy
is never exact in the region of the $\alpha$--$r_s$ plane for which
$n(k)<1$ (cf. Fig.~\ref{fig3}). For the 2DEG we find the somewhat surprising
result that for $\alpha \gtrsim 0.64$ the correlation energy is exact for densities
that violate the LO bound (cf. Fig.~\ref{fig4}). This apparent contradiction
can be resolved by remembering that $E_c$ contains kinetic contributions.
Hence, we have plotted the solid black lines in both
Figs.~\ref{fig3} and \ref{fig4}, which correspond to the critical densities
for the violation of the LO bound when the kinetic correlation is included in $E_{xc}$
\footnote{This corresponds to the DFT definition of the $xc$ energy.}.
Similarly we see that $\tilde{E}_c$, which excludes kinetic correlations,
is always in the region of the $\alpha$--$r_s$ plane where the LO bound is obeyed.

\section{Summary and conclusions}

To summarize, we have numerically solved the minimizing
momentum distributions for the power functional in
both the three- and two-dimensional homogeneous electron gas.
In particular, we have studied the ranges of validity
for the power $\alpha$ and the density parameter $r_s$ 
in terms of satisfying the Lieb-Oxford 
lower bound for $E_{xc}$ and excluding the pinned states with $n(\vk)=1$.
The results have been compared to previous limits obtained
from variational momentum distributions.

On the plane spanned by $\alpha$ and $r_s$, we have found regions 
of validity for the power functionals
at $\alpha\gtrsim 0.57$ and $r_s \gtrsim 10$ in three dimensions
and at $\alpha\gtrsim 0.60$ and $r_s \gtrsim 5$ in two dimensions.
The lower boundaries of these regions in terms of $r_s$ --
determined by the existence of pinned states -- are pushed further 
to lower values when using the numerical solutions instead
of the variational momentum distributions. However, the range
of validity corresponds to relatively low densities, significantly
lower than typical densities in, e.g., atoms, molecules, or clusters.
In two dimensions, $r_s\sim 5$ could be realized in semiconductor 
quantum-dot systems~\cite{reimann}.

We have also computed the numerically exact correlation energies
and their kinetic contributions (in terms of density-functional
theory). The exact solutions partly coincide with the regimes of 
validity, but not with the same power $\alpha$ for both quantities
$E_c$ and $t_c$. Therefore, the minimizing momentum distribution is not
the exact one even if the correlation energy is exact for a given density.

\begin{acknowledgments}
We thank Klaas Giesbertz and Paola Gori-Giorgi for helpful comments. The work was supported by the Academy of Finland through Project No. 126205 and the Nordic Innovation through its Top-Level Research Initiative Project No.~P-13053.
F. G. E. was supported by the Deutsche Forschungsgemeinschaft (DFG) through Grant No. EI 1014/1-1.
\end{acknowledgments}


\begin{thebibliography}{ll}

\bibitem{lowdin} P.-O. L\"owdin, Phys. Rev. {\bf 97}, 1474 (1955).

\bibitem{gilbert} T. L. Gilbert, Phys. Rev. B {\bf 12}, 2111 (1975).

\bibitem{dft} For a review, see, e.g., R. M. Dreizler and E. K. U. Gross,
{\it Density Functional Theory} (Springer, Berlin, 1990);
U. von Barth, Phys. Scr. {\bf T109}, 9 (2004).

\bibitem{corre1} D. R. Rohr, J. Toulouse, and K. Pernal, Phys. Rev. A {\bf 82}, 052502 (2010).

\bibitem{corre2} K. J. H. Giesbertz, E. J. Baerends, and O. V. Gritsenko, 
Phys. Rev. Lett. {\bf 101}, 033004 (2008).

\bibitem{corre3} N. N. Lathiotakis, N. Helbig, A. Rubio, and N. I. Gidopoulos,
Phys. Rev. A {\bf 90}, 032511 (2014).

\bibitem{finitetemp} T. Baldsiefen, A. Cangi, and E. K. U. Gross,
Phys. Rev. A {\bf 92}, 052514 (2015).

\bibitem{excitations} K. Pernal, J. Chem. Phys. {\bf 136}, 184105 (2012).

\bibitem{mott1} S. Sharma, J. K. Dewhurst, S. Shallcross, and E. K. U. Gross,
Phys. Rev. Lett. {\bf 110}, 116403 (2013).

\bibitem{mott2} Y. Shinohara, S. Sharma, S. Shallcross, N. N. Lathiotakis, and E. K. U. Gross, J. Chem. Theor. Comp. {\bf 11}, 4895 (2015).

\bibitem{sharma} S. Sharma,  J. K. Dewhurst, N. N. Lathiotakis, and 
E. K. U. Gross, Phys. Rev. B {\bf 78}, 201103(R) (2008).

\bibitem{nek1}  N. N. Lathiotakis, N. Helbig, and  
E. K. U. Gross, Phys. Rev. B {\bf 75}, 195120 (2007).

\bibitem{nek2} N. N. Lathiotakis, S. Sharma, J. K. Dewhurst, F. G. Eich,
M. A. L. Marques, and E. K. U. Gross, Phys. Rev. A {\bf 79}, 040501(R) (2009).

\bibitem{muller} A. M. K. M\"uller, Phys. Lett. A {\bf 105}, 446 (1984).

\bibitem{harju} E. T\"ol\"o and A. Harju, Phys. Rev. B {\bf 81}, 075321 (2010).

\bibitem{levy} M. Levy, {\em Density Matrices and Density Functionals},
(Reidel, Dordrecht, 1987).

\bibitem{kasia} J. Cioslowski and K. Pernal, 
J. Chem. Phys. {\bf 111}, 3396 (1999).

\bibitem{kasia2} J. Cioslowski and K. Pernal, 
Phys. Rev. A {\bf 61}, 034503 (2000).

\bibitem{antti} A. Putaja and E. R\"as\"anen, Phys. Rev. B {\bf 84}, 035104 (2011).





\bibitem{lo} E. H. Lieb, Phys. Lett. {\bf 70A}, 444 (1979); 
E. H. Lieb and S. Oxford, Int. J. Quantum Chem. {\bf 19}, 427 (1981).

\bibitem{ourbound}  E. R\"as\"anen, S. Pittalis, K. Capelle, and C. R. Proetto,
Phys. Rev. Lett. {\bf 102}, 206406 (2009).

\bibitem{kimball} J. C. Kimball, J. Phys. A {\bf 8}, 1513 (1975).

\bibitem{klaas1} K. J. H. Giesbertz and R. van Leeuwen, J. Chem. Phys. {\bf 139}, 104109 (2013).

\bibitem{lieb2015} M. Lewin and E. H. Lieb, Phys. Rev. A {\bf 91}, 022507 (2015).
\bibitem{paola2016} M. Seidl, S. Vuckovic, and P. Gori-Giorgi, Mol. Phys. (2016); doi:10.1080/00268976.2015.1136440.

\bibitem{lsy} E. H. Lieb, J. P. Solovej, and J. Yngvason, 
Phys. Rev. B {\bf 51}, 10646 (1995).

\bibitem{baldsiefen} T. Baldsiefen and E. K. U. Gross,
Comput. Theor. Chem. {\bf 1003}, 114 (2013).

\bibitem{VignaleBook} G. F. Giuliani and G. Vignale,
{\em Quantum Theory of the Electron Liquid},
(Cambridge University Press, 2005).

\bibitem{OB} G. Ortiz, M. Harris, and P. Ballone,
Phys. Rev. Lett. {\bf 82}, 5317 (1999).

\bibitem{AMGB} C. Attaccalite, S. Moroni, P. Gori-Giorgi, and G. B. Bachelet,
Phys. Rev. Lett. {\bf 88}, 256601 (2002).

\bibitem{Takada} Y. Takada and H. Yasuhara,
Phys. Rev. B {\bf 44}, 7879 (1991).

\bibitem{reimann} S. M. Reimann and M. Manninen, 
Rev. Mod. Phys. {\bf 74}, 1283 (2002).

\end{thebibliography}
\end{document}